\documentclass[reprint,superscriptaddress,a4paper,aps,showpacs,longbibliography]{revtex4-1}
\usepackage{graphicx}
\graphicspath{{./},{/home/user/fig/publi/muon/}}
\usepackage{amsfonts}
\usepackage{amsbsy}
\usepackage{graphicx}
\begin{document}

\title{Magnetic structure of the MnGe helimagnet and representation analysis}

\author{A. Yaouanc}
\author{P. Dalmas de R\'eotier}
\affiliation{Universit\'e Grenoble Alpes, INAC-PHELIQS, F-38000 Grenoble, France}
\affiliation{CEA, INAC-PHELIQS, F-38000 Grenoble, France}
\author{A. Maisuradze}
\affiliation{Department of Physics, Tbilisi State University, Chavchavadze 3, 
GE-0128 Tbilisi, Georgia}
\author{B. Roessli}
\affiliation{Laboratory for Neutron Scattering and Imaging, Paul Scherrer Institute, 5232 Villigen-PSI, Switzerland}
\date{\today}

\begin{abstract}

In the light of recent results obtained for the prototype helimagnet MnSi we examine the possible magnetic structures of compounds of the same family, consistent with the crystal symmetries when the magnetic propagation vector is parallel to the [001] axis. The analysis of a published muon spin rotation spectrum recorded in MnGe [Phys. Rev. B {\bf 93}, 174405 (2016)] shows no deviation from the canonical helimagnetic structure, unlike in MnSi. This qualitative difference calls for further theoretical works on chiral magnets.

\end{abstract}


\maketitle

\section{Introduction}
\label{Intro}

Noncollinear magnets are nowadays of special interest. On the one hand, they give the possibility of coupling electric and magnetic degrees of freedom as in so-called type-II multiferroic materials. The coupling may arise from the inverse Dzyaloshinskii-Moriya (DM) interaction \cite{Katsura05}. On the other hand, under a moderate applied magnetic field a peculiar spin texture might appear in a small pocket of the temperature-field phase diagram, namely a skyrmion lattice \cite{Muhlbauer09a}. A skyrmion is a magnetic topological knot.

In cubic B20 compounds (space group P2$_1$3) such as MnSi, a zero-field helical spin order is expected due to the competition of ferromagnetic and DM exchange interactions, while weaker interactions determine the characteristics of the helical structure such as the direction of the magnetic propagation wavevector ${\bf k}$ \cite{Bak80,Nakanishi80}. For example, ${\bf k}$ is found parallel to the diagonal of the cubic crystal structure for MnSi and to run along the cube edge for MnGe \cite{Makarova12} or Cu$_2$OSeO$_3$ \cite{Adams12,Seki12a}. Until recently the magnetic structure of these magnets has been described in the following way: considering an atomic plane perpendicular to ${\bf k}$, the magnetic moments are confined to some direction of that plane and ferromagnetically coupled. In subsequent atomic planes, the direction of the moments is rotated around ${\bf k}$ so that the magnetic moments describe a helix of pitch $2\pi/k$. This situation prevails in zero and small external fields. By application of a magnetic field ${\bf B}_{\rm ext}$ of sufficient strength, the magnetic structure becomes conical (except in the aforementioned skyrmion phase): the propagation vector aligns along ${\bf B}_{\rm ext}$ and the magnetic moments in the conical structure have two components. The first is perpendicular to ${\bf B}_{\rm ext}$ and corresponds to the helical component already described. The second is a uniform component aligned along ${\bf B}_{\rm ext}$.

Deviations from these regular helical and conical magnetic structures have been recently evidenced for MnSi. These results have been obtained from a detailed analysis of muon spin rotation ($\mu$SR) data while constraining the symmetries of the magnetic phases to be consistent with the crystal symmetry. In zero field, for which ${\bf k}\parallel \langle 111\rangle$, it is found that the phase of a given magnetic moment at vector position ${\bf r}$ in the crystal is not solely given by the scalar product ${\bf k}\cdot{\bf r}$. An additional phase shift differentiates the Mn positions for which the local three-fold axis is parallel to ${\bf k}$ and the others  \cite{Dalmas16}. This phase shift is even larger in the conical phase when ${\bf B}_{\rm ext}\parallel [111]$. In the case where ${\bf B}_{\rm ext}\parallel [001]$, not only may a phase shift be present between certain magnetic moments, but also the helical component lies in a plane which is not perpendicular to ${\bf B}_{\rm ext}$ \cite{Dalmas17}.

At this stage, a question arises: is MnSi the only of its kind to exhibit deviations from the regular helimagnetic structure? As mentioned elsewhere \cite{Dalmas16} these deviations are difficult to unravel in traditional neutron scattering experiments since they imply very small variations of the magnetic structure factor. On the other hand a local probe sitting at an interstitial crystallographic site like the muon is more adapted to detect them. A relevant answer to the raised question is therefore to consider the results of $\mu$SR experiments in the ordered phase of helimagnets.

While Cu$_2$OSeO$_3$ belongs to this family \cite{Maisuradze11,Lancaster15}, its structure is complicated by the presence of two Cu crystallographic sites. This is not so for MnGe which is isostructural to MnSi. Recently this B20 metal has caught the attention of experimentalists because of its giant topological Hall effect \cite{Kanazawa11}. It has been argued \cite{Makarova12} to order magnetically at low temperature in a helical structure with ${\bf k}$ parallel to $[001]$ and to exhibit a cubic lattice of skyrmions \cite{Tanigaki15}. An important specificity of MnGe is its short helix pitch, which could emphasize the deviations from the regular helix already found in MnSi. In this paper, using representation analysis \cite{Bertaut81}, we derive the possible magnetic structure of a B20 magnet approaching the regular helimagnetic structure for ${\bf k} \parallel \langle 001\rangle$ and compatible with the crystal symmetry. Then the analysis of a zero-field (ZF) muon spin rotation ($\mu$SR) spectrum recorded at 10~K by Martin and coworkers \cite{Martin16} allows us to conclude on the absence of deviation within the accuracy of the parameters.

The organization of this paper is as follows.  Section~\ref{Survey} gives a survey of the physical properties of MnGe of interest for our work.  In Sec.~\ref{Group_summary} we summarize the predictions of representation analysis as applied to the determination of the possible magnetic structures of MnGe. In Sec.~\ref{result}, the analysis of a published ZF $\mu$SR spectrum is described. We finally present our conclusions in Sec.~\ref{Conclusions}.  Representation analysis for MnGe is exposed thoroughly in Appendix~\ref{Group_details}. The expression of the polarization function used for the maximum-entropy--reverse-Monte-Carlo computation is given in Appendix~\ref{ME/RMC}.

\section{Basic physical properties of M\lowercase{n}G\lowercase{e}}
\label{Survey} 

In its paramagnetic phase, the metallic compound MnGe crystallizes with the cubic space group P2$_1$3. It magnetically orders at $T_{\rm c} = 170 \, (5)$~K \cite{Kanazawa11}. Neutron powder diffraction indicates the compound to be orthorhombic in its ordered state with space group P$2_12_12_1$ \cite{Makarova12}.  The three measured lattice parameters --- $a_{\rm lat} = 4.7806 \, (30)$,  $b_{\rm lat} = 4.7805 \, (29)$, and $c_{\rm lat} = 4.7939 \, (10)$~\AA; values obtained at 6~K \cite{Makarova12} --- are nearly equal. The Mn ions occupy a $4a$ position in Wyckoff's notation. Their coordinates depend on three parameters, i.e.\ $x_{\rm Mn} = 0.142 \, (12)$, $y_{\rm Mn} = 0.131 \, (16)$, and $z_{\rm Mn} = 0.136 \, (11)$ at 6~K. Since the three lattice parameters and the three position parameters listed above are almost equal, MnGe is close to the cubic P$2_13$ space group below $T_{\rm c}$. For reference, Table~\ref{table_coordinates} lists the positions of the four Mn atoms in the unit cell. The magnetic propagation wavevector ${\bf k}$ is directed along the $c$ axis of the orthorhombic structure. Its modulus saturates to $k =2.19 \, (5)$~nm$^{-1}$ below $T_{\rm com} = 30$~K  \cite{Makarova12}. This corresponds to a helix period of $2 \pi/k \simeq 2.9$~nm.  Like in MnSi \cite{Grigoriev10}, the handedness of the MnGe crystalline structure determines the chirality of the magnetic structure \cite{Grigoriev13}. The Mn magnetic moment $m$ at low temperature in MnGe is relatively large. From neutron diffraction $m = 2.3 \, (5) \, \mu_{\rm B}$ \cite{Makarova12}, latterly refined to $m = 1.83 \, (15) \, \mu_{\rm B}$ \cite{Deutsch14}. On the other hand, bulk magnetic measurements performed at 2.5~K under 14~T for a powder sample lead to only $m \approx 1.65 \, \mu_{\rm B}$ \cite{Deutsch14a}. Remarkably, the measured moment does not saturate even under this large field.

\begin{table}
\caption{Coordinates of the $4a$ equivalent positions for Mn in the crystal unit cell of MnGe at low temperature where the crystallographic space group is P2$_1$2$_1$2$_1$. The positions are labeled by $\gamma$. For reference, the last column gives the three-digit numerical coordinates of these positions in the origin cell, i.e.\ the ${\bf d}_\gamma$ vector coordinates. The numerical data have been determined from measurements performed at 6~K \cite{Makarova12}. The values refer to a right handedness. The coordinates for Mn in a crystal of the left handedness are obtained by taking their complement to 1. As an example they are (0.858, 0.869, 0.864) for $\gamma$ = I.}
\label{table_coordinates}
\begin{tabular}{l@{\hskip 3mm}l@{\hskip 3mm}l@{\hskip 3mm}l}
\hline\hline
Position & $\gamma$ & coordinates & corresponding\cr
     & & &  coordinates in \cr
     & & & origin cell \cr\hline
$4{\rm a}$   & I      & $(x_{\rm Mn},y_{\rm Mn},z_{\rm Mn})$ & (0.142, 0.131, 0.136)\cr
             & II     & $(x_{\rm Mn}+\frac{1}{2}, \bar{y}_{\rm Mn}+\frac{1}{2}, \bar{z}_{\rm Mn})$ & (0.642, 0.369, 0.864)\cr
             & III    & $(\bar{x}_{\rm Mn}, y_{\rm Mn}+\frac{1}{2}, \bar{z}_{\rm Mn}+\frac{1}{2})$ & (0.858, 0.631, 0.364) \cr
             & IV     & $(\bar{x}_{\rm Mn}+\frac{1}{2}, \bar{y}_{\rm Mn}, z_{\rm Mn}+\frac{1}{2})$ & (0.358, 0.869, 0.636)\cr
\hline\hline
\end{tabular}
\end{table}

\section{Symmetry analysis of the magnetic structures at the M\lowercase{n} sites}
\label{Group_summary} 

Magnetic structures compatible with a crystal symmetry can be inferred with the help of representation analysis, i.e.\ Bertaut's theory; see Ref.~\onlinecite{Bertaut81} and references therein. We point out that we are dealing with a geometrical rather than a thermodynamical problem. So the order of the magnetic phase transition does not interfere with our search of the possible magnetic structures.

While a single K-domain exists in the orthorhombic structure for ${\bf k}$ parallel to the $c$ axis, two spin-domains are present. However, as explained elsewhere \cite{Dalmas16}, they have no influence on a $\mu$SR field distribution when the magnetic structure is incommensurate. 

\begin{figure}
\centering
   \includegraphics[width=0.25\textwidth]{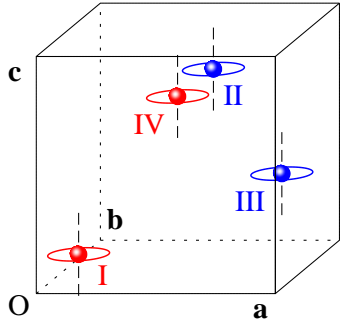}
\caption{(Color online) Schematic illustration of the magnetic structure of MnGe. The magnetic moments at each of the four Mn sites in the unit cell lie in planes perpendicular to ${\bf k}$ = $[0,0,k]$. The relative phase of the moments at sites I and IV is given by the variation of the scalar product ${\bf k}\cdot{\bf r}$ between these two sites. The same result holds for sites II and III. However representation analysis does not enforce any relation between the phases at sites I or IV in the one hand and sites II or III in the other hand.}
\label{magnetic_structure}
\end{figure}

The spin structures compatible in general with the crystal structure are determined in Appendix~\ref{Group_details}.  Since the ground state is believed to have a helical magnetic state, here we search for a state closely related to it. Magnetic moments are assumed to have an equal modulus and to rotate in planes normal to ${\bf k}$. Denoting ${\bf m}_{i + d_\gamma}$ the magnetic moment at lattice position ${\bf i} + {\bf d}_\gamma$, the following generic formula is found to hold:
\begin{eqnarray}
{\bf m}_{i + d_\gamma}  =   
m \left[  \cos \left({\bf k}\cdot{\bf i}\right) \tilde{\bf a}_{d_\gamma}
  - \sin \left({\bf k}\cdot{\bf i}\right) \tilde{\bf b}_{d_\gamma} \right].
 \label{Group_theory_application_structure_0}
\end{eqnarray}
The sign ``$-$'' in front of the sine function stands for the left-handed magnetic chirality observed for the right handedness of the structure \cite{Grigoriev13}. For a right-handed magnetic chirality, this is a ``+'' sign. In Eq.~\ref{Group_theory_application_structure_0}, $\gamma \in \{{\rm I}, {\rm II}, {\rm III}, {\rm IV} \}$ labels the four Mn positions, ${\bf i}$ denotes the position of a unit cell in the crystal and ${\bf d}_\gamma$ specifies the position of atom $\gamma$ in the unit cell. Finally $(\tilde{\bf a}_{d_\gamma}, \tilde{\bf b}_{d_\gamma}, {\bf k}/k)$ denotes a right-handed orthonormal reference frame. Equation~\ref{moment_gamma_2} enforces a certain relationship between the orientations of vectors $\tilde{\bf a}_{d_\gamma}$ and $\tilde{\bf b}_{d_\gamma}$ for different $\gamma$'s.  
In the following, we find it convenient to express Eq.~\ref{Group_theory_application_structure_0} in terms of the absolute position ${\bf i} + {\bf d}_\gamma$ in the crystal. Setting $\alpha_{i,d_\gamma} \equiv {\bf k}\cdot ({\bf i + d_\gamma})$ and introducing a phase $\beta_{d_\gamma}$ independent of $i$, we arrive at
\begin{eqnarray}
{\bf m}_{i + d_\gamma}  =   
 m \left[  \cos \left(\alpha_{i,d_\gamma} 
+ \beta_{d_\gamma} \right) {\bf a}_{d_\gamma} 
-  \sin \left(\alpha_{i,d_\gamma}
+ \beta_{d_\gamma} \right) {\bf b}_{d_\gamma} \right],\cr
 & & \label{Group_theory_application_structure_1}
\end{eqnarray}
with
\begin{eqnarray}
{\bf a}_{d_\gamma} & = & (1,0,0) \ \ {\rm and} \ \ {\bf b}_{d_\gamma} = (0,1,0).
\label{Group_theory_application_structure_2}
\end{eqnarray}
Note that $ {\bf a}_{d_\gamma}$ and $ {\bf b}_{d_\gamma}$ are actually independent of $\gamma$. 
Table~\ref{phase} 
\begin{table}
\caption{Phases for the Mn magnetic moments in MnGe as inferred from representation analysis. }
\label{phase}
\begin{tabular}{lcccc}
\hline \hline
$\gamma$         & ${\rm I}$ & ${\rm II}$ & ${\rm III}$ & ${\rm IV}$ \cr \hline
$\beta_{d_\gamma}$ & 0         & $\phi$  & $\phi$   & 0 \cr
\hline \hline
\end{tabular}
\end{table}
lists the values for $\beta_{d_\gamma}$ consistent with representation analysis and Fig.~\ref{magnetic_structure} provides an illustration for the relation between the different phases. The magnetic structure being incommensurate and the field distribution at the muon site depending only on the difference between the phases at sites I and IV in the one hand and II and III on the other hand, we arbitrarily fix $\beta_{d_\gamma}$ to 0 for $\gamma$ = I and IV. In Eq.~\ref{Group_theory_application_structure_1} we have two free parameters: the phase $\phi$ defined in Table~\ref{phase} and the magnetic moment modulus $m$. The regular helical structure corresponds to a vanishing phase $\phi$.

\section{Analysis of a ZF $\mu$SR spectrum of M\lowercase{n}G\lowercase{e} at 10~K based on symmetry}
\label{result}

\subsection{The polarization function}\label{Analysis_pol_func}

The computation of the $\mu$SR polarization function $P_Z(t)$ follows the method introduced recently \cite{Dalmas16}. Basically, the polarization function associated with muons stopped at position $s_\eta$ and for an orientation $o$ of the crystal relative to the laboratory reference frame, i.e.\ $P_{Z,s_\eta, o}(t)$, is first evaluated. For this purpose the spontaneous field at the muon site is computed. It comprises the dipole field arising from the localized magnetic moments in the crystal and the contact field associated with the electron density at the muon position. The fields are conveniently computed using Fourier transforms and therefore we use the Fourier transform of ${\bf m}_{i + d_\gamma}$. The finite coherence length $\xi$ of the magnetic structure is explicitly taken into account through an integral over the wavevectors in the vicinity of $\pm{\bf k}$ \cite{Dalmas16}. The spin-lattice relaxation channel is characterized by the relaxation rate $\lambda_Z$ and the damping of the muon precession arising from the $^{55}$Mn nuclear dipoles is described by the parameter $\Delta_{\rm N}$ which is the root-mean-square of the nuclear field distribution.
Notice that the contribution of the Ge nuclei is negligible. The effect of the spin-spin relaxation channel, which, in simple models, scales with the spin-lattice relaxation has been discarded. The model polarization function $P_Z(t)$ is obtained from an average of $P_{Z,s_\eta, {\rm o}}(t)$ over the muon positions and the orientations since the available data concern a polycrystal.

\subsection{Results}\label{results}

\begin{figure}
\centering
   \includegraphics[width=0.45\textwidth]{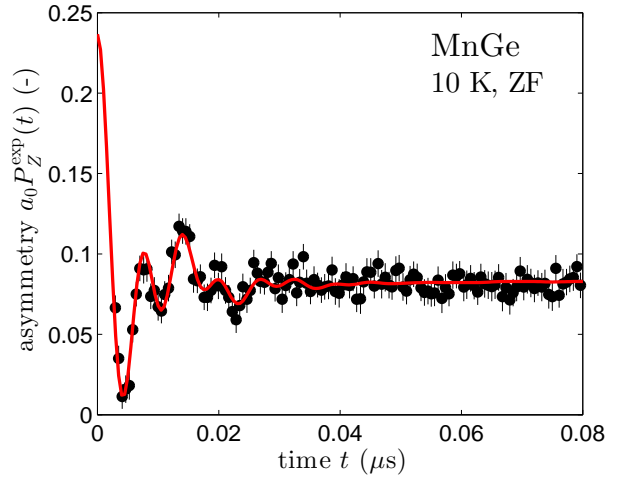}
\caption{(Color online) 
  ZF $\mu$SR asymmetry spectrum recorded for a polycrystal of MnGe at 10~K. The data are reproduced from Ref.~\cite{Martin16}. The solid line represents the fit discussed in the main text.
  Notice the short-time range over which the spectrum is displayed. This is due to the relatively large frequencies detected and their associated strong damping.}
\label{Spectrum}
\end{figure}

The ZF $\mu$SR asymmetry spectrum $a_0P^{\rm exp}_z(t)$ published in Ref.~\onlinecite{Martin16} is reproduced in Fig.~\ref{Spectrum}. 
It is analyzed with a two-component model:
\begin{equation}
 a_0P^{\rm exp}_Z(t) = a_0 \left[ (1-f_{\rm bg})P_Z^{\rm main}(t) + f_{\rm bg}\right],
 \label{eq:AsyModel:LF}
\end{equation}
where $a_0$ is the initial $\mu$SR asymmetry and $f_{\rm bg}$ the fraction of muons stopped outside of the sample, for which the relaxation is negligible. We shall find that a description of the spectrum requires to take into account a secondary phase contribution:
\begin{equation}
 P^{\rm main}_Z(t) = (1 - f_{\rm imp}) P_Z^{\rm MnGe}(t) + f_{\rm imp} P_Z^{\rm imp}(t),
 \label{pol1}
\end{equation}
where $P_Z^{\rm MnGe}(t)$ is the polarization function discussed in Sec.~\ref{Analysis_pol_func}, i.e. $P_Z^{\rm MnGe}(t) = P_Z(t)$, and
\begin{equation}
 P_Z^{\rm imp}(t) = {1 \over 3} + {2 \over 3}
\left [  \left (1 - \gamma_\mu \Delta_{\rm L} t \right )\exp \right (-\gamma_\mu \Delta_{\rm L}  t\left )  \right ],
 \label{pol2}
\end{equation}
models an impurity phase. $P_Z^{\rm imp}(t)$ as given here is typical for a diluted disordered magnetic system \cite{Yaouanc11}. It corresponds to a squared-Lorentzian distribution for the modulus of the field at the muon site with a half-width at half-maximum equal to $ \Delta_{\rm L}\sqrt{\sqrt{2} - 1} = 0.644\, \Delta_{\rm L}$. 

\begin{figure}
\centering
   \includegraphics[width=0.45\textwidth]{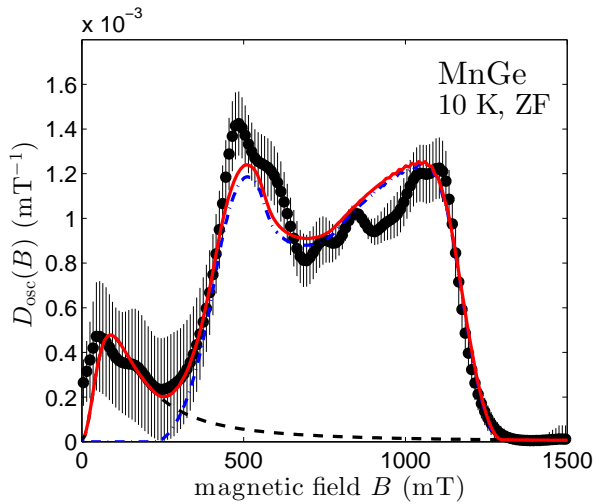}
\caption{(Color online) 
  Field distribution associated with the spectrum displayed in Fig.~\ref{Spectrum}. For each data point we have computed the experimental uncertainty (one standard deviation). The red solid line results from the fit to the asymmetry spectrum (Fig.~\ref{Spectrum}), i.e.\ it is not a fit to $D_{\rm osc}(B)$, as explained in the main text.
The dashed-dotted blue line displays the contribution of MnGe to the field distribution and the lower dashed black line accounts for a magnetically disordered secondary phase contained in the sample. The relatively large error bars on the distribution are explained by the correlations existing between neighbor points.
}
\label{Fourier}
\end{figure}

While only $a_0P^{\rm exp}_z(t)$ is fitted, it is useful to consider the field distribution associated with the oscillating part of the spectrum, namely $D_{\rm osc}(B)$; see Appendix~\ref{ME/RMC} for its definition. This quantity is computed from the asymmetry spectrum of Fig.~\ref{Spectrum} using the maximum entropy (ME) principle combined with the reverse Monte Carlo (RMC) algorithm; see Ref.~\onlinecite{Dalmas14} and appendix of Ref.~\onlinecite{Dalmas16a} for an exposition of the ME--RMC method. Its main advantages over an inverse Fourier transform are twofold. (i) Original data uncertainties are taken into account, leading to the reduced noise relative to the output of the conventional Fourier transform. (ii) Error bars on the field distribution are estimated. Figure~\ref{Fourier} displays the result. We observe a relatively sharp peak at $\approx 0.5$~T and a second wider skewed peak at $\approx 1.1$~T. In addition, a third weak maximum is present at low field. The distribution is relatively sharp-cut at high field. 

Before discussing the result of the fit, a few comments on the computation are in order.
Since $\Delta_{\rm N}$ is expected to be approximately the same for MnGe and MnSi, we set $\Delta_{\rm N} = 1.11$~mT as found for MnSi \cite{Dalmas16}. Because of the strong damping of the spontaneous oscillations --- see Fig.~\ref{Spectrum} --- the precise $\Delta_{\rm N}$ value does not really influence our result. As to the determination of $\xi \equiv 1/\kappa$, an integration over the reciprocal lattice has to be performed. Here we ensured the $({\bf a}_{d_\gamma},{\bf b}_{d_\gamma} )$ plane to remain perpendicular to the wavevector at each step of the integration. While forgetting this fact has a negligible effect on the final result for a compound with a large $\xi$ such as MnSi, this leads to a flawed estimate of $\xi$ if it is relatively short as for MnGe. Concerning the wavevector integration range, we found spheres of radius $10\,\kappa$ centered around $\pm{\bf k}$ to be sufficient for the evaluation of $\kappa$. As mentioned above, in principle an orientation average has to be performed since the sample is polycrystalline. However, we found numerically that the polarization function is independent of the orientation of the crystal in the muon beam and therefore no average over the orientations is needed. This makes it possible to significantly reduce the numerical effort. Unlike in Ref.~\onlinecite{Dalmas16} and because of the large mean magnetic field at the muon site and its important standard deviation --- see Fig.~\ref{Fourier} --- we took the time resolution of the spectrometer into account \cite{note_resolution}. Finally, since simulated spectra for the orthorhombic crystal structure are virtually identical to those computed in a cubic crystal structure, the final fit was performed assuming the cubic crystal structure. The unique parameter describing the muon site in space group P2$_1$3 is defined as $x_\mu$.

In a first instance fits were performed with the model previously described excluding the integration over the wavevector which requires more intensive calculations. Satisfactory results were found with $m$ and $x_\mu$ in the expected range. The angle $\phi$ was found consistent with 0, within error bars. However a relatively strong correlation was found in the values for these parameters. Therefore in the final fit including the effect of the finite value for $\xi$, $m$ was set to the value obtained from neutron diffraction ($m$ = 1.83\,$\mu_{\rm B}$) and $\phi$ to 0. The result is shown as a solid line in Fig.~\ref{Spectrum}. It is excellent. The solid line in Fig.~\ref{Fourier} is a byproduct of the asymmetry spectrum fit.

The fit parameters are as follows: $\xi$ = 9\,(1)\,nm, $x_\mu$ = 0.54\,(1), and the parameter describing the contact interaction between the Mn moments and the muon \cite{Dalmas16,Yaouanc11} is $r_\mu H/4\pi = -1.47\,(1)$. The resulting Fermi contact field is ${\bf B}_{\rm con} = \frac{\mu_0}{v_0}\frac{r_\mu H}{4\pi}\overline{\bf m}$, where $\mu_0$ is the permeability of free space, $v_0$ is the volume per Mn ion and $\overline{\bf m}$ is the vector average of the Mn magnetic moments in interaction with the muon. It is antiparallel to the local magnetization and its magnitude is 1.10\,(1)\,T in agreement with Ref.~\onlinecite{Martin16}. The initial asymmetry $a_0=0.245$ and the spin-lattice relaxation rate $\lambda_Z=0.05~\mu$s$^{-1}$ were taken from Ref.~\onlinecite{Martin16}. In addition, $f_{\rm bg} = 0.0011 \, (3)$. So the background is negligibly small. For the impurity phase we get $f_{\rm imp} = 0.13 \, (2)$ and $\Delta_{\rm L} = 88 \, (14)$~mT.

\subsection{Discussion}\label{discussion}

Another fit to the data was performed assuming the magnetic structure to be of the other chirality (see below Eq.~\ref{Group_theory_application_structure_0}), and therefore a crystal with the alternate handedness. An equivalent agreement with the data is found with the same value for all the parameters except the muon position parameters $x_\mu$. Still the two $x_\mu$ parameters are found linked by the correspondence $x_\mu \longleftrightarrow 1-x_\mu$, as expected from the difference in the structure handednesses. Therefore we cannot conclude on the handedness of the sample, as was already the case for MnSi. Considering the left-handed magnetic chirality results, we note that the position parameter $x_\mu = 0.54\,(1)$ for the muon in MnGe is close to that of MnSi ($x_\mu=0.532$; Refs.~\onlinecite{Amato14,Dalmas16}). This is in accord with the {\em ab initio} computations which predict a value for $x_\mu$ slightly larger in MnGe than in MnSi, namely 0.554 vs 0.542 \cite{Bonfa15,Martin16}.

The free phase $\phi$ allowed by representation analysis is found to be 0. This is unlike MnSi for which deviations from the regular helimagnetic structure were found both in zero-field and in applied field as far as the helical component is concerned \cite{Dalmas16,Dalmas17}. 
The magnetic disorder appears quite strong as reflected by the relatively small $\xi$ value. This is in line with the small number of visible oscillations in the experimental spectrum (Fig.~\ref{Spectrum}) and at variance to the MnSi case. No such small coherence length was signaled in the MnGe neutron diffraction work \cite{Makarova12}, but the strong damping of the $\mu$SR oscillations was associated with the disorder inherent to the magnetic structure \cite{Martin16}. Note that with our model, the parameter $\xi$ accounts in fact for the effects of both magnetic disorder and spin-spin relaxation. If this relaxation is not negligible, our $\xi$ value is underestimated. The hyperfine coupling parameter $r_\mu H/4\pi$ is negative as usual. It is even higher in absolute value than for MnSi for which the coupling was already higher than for other metals. 

The spectral weight in the interval 300 -- 1200~mT of the field distribution (Fig.~\ref{Fourier}) is ascribed to the MnGe magnetic phase. The remaining part, i.e.\ essentially in the field range below 300~mT is attributed to an impurity phase of approximately $13  \% $ volume fraction. Since it is not observed in diffraction techniques \cite{Makarova12} and because of the $P_Z^{\rm imp}(t)$ functional shape, we suggest the phase to be amorphous. MnGe being synthesized at high pressure, it may not be surprising to find such a phase. Besides, two foreign phases amounting to a total of less than $5 \, \%$ and assigned to Mn$_{11}$Ge$_8$ and Mn$_2$O$_3$ were identified by x-ray diffraction \cite{Makarova12}. Since both phases are magnetic at 10~K, they could contribute to the peak found around 0.5~T in the field distribution (Fig.~\ref{Fourier}) which is not fully accounted for by our model.

Because of the strong damping of the oscillations, the amount of information which can be extracted from the $\mu$SR spectrum is rather restricted. Hence, we have not attempted to test the possibility --- allowed by representation analysis --- for the moments to rotate in planes not normal to ${\bf k}$. It would require to introduce at least two additional free angles. Using the magnetic structure information contained in Eq.~\ref{sol_gamma_1}, it would be worthwhile to do it through a combined fit of the $\mu$SR spectrum and a high statistics neutron diffraction pattern.

Finally, we note that the magnetic fluctuations detected far below $T_{\rm c}$ in MnGe \cite{Martin16} follow the same trend as in MnSi; see e.g. Refs.~\onlinecite{Takigawa80,Yaouanc05}.

\section{Conclusions}\label{Conclusions}

In this work, we have examined the magnetic structures compatible with the symmetry for systems crystallizing in the B20 phase and with a magnetic propagation vector ${\bf k}$ parallel to an edge of the cubic crystal structure. Beyond the well-known helimagnetic phase, a more complex structure is possible and characterized by a dephasing between certain magnetic moments in the unit cell. The moments affected by the dephasing differ from those found when ${\bf k}$ is along a diagonal of the cube. 
The result derived from representation analysis applies not only for the cubic crystal structure but also for the orthorhombic structure proposed from a neutron scattering study of MnGe in its magnetic phase.

The present study uses the novel framework developed in Ref.~\onlinecite{Dalmas16} for a detailed refinement of subtle spin textures from $\mu$SR data. A recently published MnGe ZF-$\mu$SR spectrum is found consistent with a regular helimagnetic structure. The deviation from this structure found for MnSi is therefore not a generic feature of the B20 phase magnets. This illustrates the subtleties in the interaction interplay in these systems which were so far believed to be qualitatively similar. It is therefore worthwhile to revisit the assumed magnetic structures of other helimagnets. For this purpose a combined analysis of neutron diffraction patterns and $\mu$SR spectra would be most powerful. We expect a deeper insight into the interactions in this series of compounds which is considered as the playground for systems with potential applications in information storage.

\begin{acknowledgments}
We thank an anonymous referee for a relevant remark on representation analysis.
\end{acknowledgments}

\appendix

\section{Determination of the possible magnetic structures from representation analysis}
\label{Group_details} 

We recall that ${\bf k} = [0,0,k]$. The following results are derived for Wyckoff position $4a$ in space group P2$_1$2$_1$2$_1$. Interestingly, the same results are obtained for position $4a$ in space group P2$_1$3 for which the three coordinates of position $\gamma$ = I are equal.

The little group of the propagation vector contains the symmetry elements $L^{\bf k} = \{1|(0,0,0),2_{x00}|(0.5,0.5,0)\}$, as expressed with Seitz' notation. However, there is a symmetry element in the space group that transforms ${\bf k} \to - {\bf  k}$. The structure lacking the inversion symmetry, ${\bf k}$ and $-{\bf k}$ are not equivalent.  Therefore the magnetic little group is $M^{\bf k} = L^{\bf k} + \theta g L^{\bf k}$ where $\theta$ is the time-reversal operator and $g$ is one of the symmetry elements that reverse ${\bf k}$. Here for the calculations of the irreducible co-representations (ir-coreps) of $M^{\bf k}$ we chose $g=\{2_{0y0}|(0,0.5,0.5)\}$. The ir-coreps of $M^{\bf k}$ are tabulated in Table~\ref{coreps} where we have defined
\begin{equation}
  \varepsilon = \exp(i \varphi_{\rm t}), \hspace{5mm} {\rm with} \hspace{5mm} \varphi_{\rm t} = k\,c_{\rm lat}/2.
\label{epsilon}
\end{equation}
\begin{table}
\caption{The ir-coreps of the magnetic little group for ${\bf k} = [0,0,k]$. Here $\varepsilon$ is a phase factor defined in Eq.~\ref{epsilon} and $\theta$ the time-reversal operator. }
\label{coreps}
\resizebox{\linewidth}{!}{%
\begin{tabular}{c|c|c|c|c}
\hline 
          &$ \{1|(0,0,0)\}$    &$\{2_{00z}|(0.5,0,0.5)\}$                &$\theta \{2_{0y0}|(0,0.5,0.5)\}$  &$\theta \{2_{x00}|(-0.5,0.5,0)\}$\\ \hline
$\Gamma^+_1$     & 1       &  $\varepsilon^{-1} $  &  1    &$\varepsilon $\\ \hline
$\Gamma^-_1$     & 1       &  $\varepsilon^{-1} $  & $-1$    &$-\varepsilon $\\ \hline
$\Gamma^+_2$     & 1       &  $- \varepsilon^{-1} $ &  1    & $-\varepsilon $\\ \hline
$\Gamma^-_2$     & 1       &  $- \varepsilon^{-1} $ & $-1$    &$\varepsilon $\\ \hline
\end{tabular}
}
\end{table}
The structure of MnGe contains one crystallographic site for the Mn atoms as seen from Table~\ref{table_coordinates}. Applying the symmetry elements of $M^{\bf k}$ to the Mn-atomic positions of MnGe, we find the four Mn atoms to belong to a single crystallographic orbit. The decomposition of the magnetic representation is as follows:  
\begin{eqnarray} 
\Gamma_{\rm mag} = 6 \Gamma_1^+ \oplus 6 \Gamma_1^- \oplus 6 \Gamma_2^+ \oplus 6 \Gamma_2^-.
\label{Group_details_4a_1}
\end{eqnarray}
From Eq.~\ref{Group_details_4a_1} we expect six symmetry-allowed basis vectors for each co-representation that we will denote $F_j$, with $1 \le j \le 6$, for the four ir-coreps. The $F_j$ are linear combinations of the Fourier components $S_{d_\gamma,\alpha}$ and $S^*_{{\rm d}_\gamma,\alpha}$. Here $\alpha$ denotes a Cartesian component of, for example, ${\bf S}_{d_\gamma}$ and $d_\gamma$ stands for a sublattice. Finally the components of the  magnetic moment at each crystallographic position are given by  
\begin{eqnarray}
m_{i+d_\gamma,\alpha} =S_{d_\gamma,\alpha} \exp (- i{\bf k} \cdot {\bf i} ) + {\rm c.c.}  
\label{Group_details_4a_7}
\end{eqnarray}

Real solutions for the magnetic moments are found after taking a linear combination of the basis functions of $\Gamma^+_1$ and $\Gamma^-_1$, in the one hand, or $\Gamma^+_2$ and $\Gamma^-_2$, in the other hand. For $\Gamma_1$, resulting from the sum of $\Gamma^+_1$ and $\Gamma^-_1$, the basis functions are
\begin{eqnarray}
F_1 & = & S_{d_{\rm I},x} - \varepsilon S_{d_{\rm IV},x},\cr
F_2 & = & S_{d_{\rm I},y} - \varepsilon S_{d_{\rm IV},y},\cr
F_3 & = & S_{d_{\rm I},z} + \varepsilon S_{d_{\rm IV},z},\cr
F_4 & = & S_{d_{\rm II},x} - \varepsilon^{-1} S_{d_{\rm III},x},\cr
F_5 & = & S_{d_{\rm II},y} - \varepsilon^{-1} S_{d_{\rm III},y},\cr
F_6 & = & S_{d_{\rm II},z} + \varepsilon^{-1} S_{d_{\rm III},z}.
\label{Group_details_4a_6}
\end{eqnarray}
The modes for $\Gamma_2$, resulting from the combination of $\Gamma^+_2$ and $\Gamma^-_2$ are deduced by substituting ${\bf S}_{d_{\rm IV}}$ and ${\bf S}_{d_{\rm III}}$ with $-{\bf S}_{d_{\rm IV}}$ and $-{\bf S}_{d_{\rm III}}$, respectively, in Eq.~\ref{Group_details_4a_6}.

The $F_j$ functions describe the magnetic modes compatible with space group symmetry, for the given orientation of ${\bf k}$. Therefore, if the four spins magnetically order according, for example, to $\Gamma_1$, the $F_j$ functions for $\Gamma_2$ vanish \cite{Bertaut81}. As a result we get the following relations:
\begin{eqnarray}
S_{d_{\rm I},x} & = & - \varepsilon S_{d_{\rm IV},x}   \, \, \, {\rm and } \, \, \,   S_{d_{\rm II},x} = -\varepsilon^{-1}  S_{d_{\rm III},x}, \cr
S_{d_{\rm I},y} & = & -\varepsilon S_{d_{\rm IV},y}  \, \, \, {\rm and } \, \, \,   S_{d_{\rm II},y} = -\varepsilon^{-1} S_{d_{\rm III},y}, \cr
S_{d_{\rm I},z} & = & \varepsilon S_{d_{\rm IV},z}  \, \, \, {\rm and } \, \, \,   S_{d_{\rm II},z} =  \varepsilon^{-1} S_{d_{\rm III},z}, 
\label{sol_gamma_1}
\end{eqnarray}
for $\Gamma_1$. For $\Gamma_2$ the substitutional rule enounced above applies.

It is believed that the spins rotate in planes perpendicular to ${\bf k}$. We take this result for granted and therefore set $S_{d_\gamma,z}  = 0$. Without loss of generality we can write for each irrep $S_{d_{\rm I},x} = a_{x_1} \exp(i\phi_{x_1})$, $S_{d_{\rm II},x} = a_{x_2} \exp(i\phi_{x_2})$, $S_{d_{\rm I},y} = a_{y_1} \exp(i\phi_{y_1})$ and $S_{d_{\rm II},y} = a_{y_2} \exp(i\phi_{y_2})$ where $a_{x_1}$, $a_{x_2}$, $a_{y_1}$, $a_{y_2}$, $\phi_{x_1}$, $\phi_{x_2}$, $\phi_{y_1}$, and $\phi_{y_2}$ are real numbers. Setting $u_1$ = $2a_{x_1} \cos\phi_{x_1}$, $v_1$ = $2a_{x_1} \sin \phi_{x_1}$, $u_2$ = $2a_{y_1} \cos\phi_{y_1}$, $v_2$ = $2a_{y_1} \sin\phi_{y_1}$, ${\tilde u}_1$ = $2a_{x_2} \cos\phi_{x_2}$, ${\tilde v}_1$ = $2a_{x_2} \sin \phi_{x_2}$, ${\tilde u}_2$ = $2a_{y_2} \cos\phi_{y_2}$ and ${\tilde v}_2$ = $2a_{y_2} \sin\phi_{y_2}$, and using the constraints imposed by Eq.~\ref {sol_gamma_1}, we arrive at
\begin{eqnarray}
m_{i + d_{\rm I},x} & = & u_1 \cos ({\bf k}\cdot {\bf i}) + v_1 \sin ({\bf k}\cdot {\bf i}),  \label{moment_gamma_2}\\
m_{i + d_{\rm I},y} & = & u_2 \cos ({\bf k}\cdot {\bf i}) + v_2 \sin ({\bf k}\cdot {\bf i}), \cr
m_{i + d_{\rm II},x} & = & {\tilde u}_1 \cos ({\bf k}\cdot {\bf i}) + {\tilde v}_1 \sin ({\bf k}\cdot {\bf i}), \cr
m_{i + d_{\rm II},y} & = & {\tilde u}_2 \cos ({\bf k}\cdot {\bf i}) + {\tilde v}_2 \sin ({\bf k}\cdot {\bf i}), \cr
m_{i + d_{\rm III},x} & = &  {\tilde u}_1 \cos ({\bf k}\cdot {\bf i}-\varphi_{\rm t}) + {\tilde v}_1 \sin( {\bf k}\cdot {\bf i}-\varphi_{\rm t}), \cr
m_{i + d_{\rm III},y} & = &  {\tilde u}_2 \cos ({\bf k}\cdot {\bf i}-\varphi_{\rm t}) + {\tilde v}_2 \sin( {\bf k}\cdot {\bf i}-\varphi_{\rm t}), \cr
m_{i + d_{\rm IV},x} & = &  u_1 \cos ({\bf k}\cdot {\bf i}+\varphi_{\rm t}) + v_1 \sin( {\bf k}\cdot {\bf i}+\varphi_{\rm t}), \cr
m_{i + d_{\rm IV},y} & = &  u_2 \cos ({\bf k}\cdot {\bf i}+\varphi_{\rm t}) + v_2 \sin( {\bf k}\cdot {\bf i}+\varphi_{\rm t}), \nonumber
\end{eqnarray}
for $\Gamma_2$.

The magnetic structure depends on eight free parameters for the eight magnetic moment components. From Eq.~\ref{moment_gamma_2}, recalling the definition of $\varphi_{\rm t}$ (Eq.~\ref{epsilon}), we derive the solution given in Eq.~\ref{Group_theory_application_structure_1} of the main text, with parameters defined in Eq.~\ref{Group_theory_application_structure_2} and Table~\ref{phase}. The case of the regular helimagnet corresponds to $\phi$ = 0. We have not retained the $\Gamma_1$ irrep because it leads to a magnetic structure even more distant from the regular helical structure.

\section{Expression of the polarization function in terms of $D_{\rm osc}$}
\label{ME/RMC} 

The field distribution $D_{\rm osc}(B)$ associated with the oscillating part of the spectrum was computed using the ME--RMC method with $P_Z^{\rm main} (t)$ in Eq.~\ref{eq:AsyModel:LF} given by the following expression:
\begin{eqnarray}
P_Z^{\rm main} (t) & = & {1 \over 3}\exp\left ( - \lambda_Z t \right ) \label{ME}\\
& + & {2 \over 3} \int
( 1 - {\mathcal T} B ) D_{\rm osc} (B) \cos \left (\gamma_\mu B t  \right ) {\rm d} B.
\nonumber
\end{eqnarray}
The parameter ${\mathcal T}$ accounts for the spectrometer finite resolution; see also main text and Ref.~\onlinecite{note_resolution}.

\bibliography{reference,MnGe_note}

\end{document}